\documentclass[9pt,twocolumn,twoside]{opticajnl}
\journal{opticajournal} 

\setboolean{shortarticle}{true}

\usepackage{amsmath}
\usepackage{lineno}
\usepackage{mathrsfs}

\title{Spectral tailoring of Raman soliton generation via a dispersion-managed fibre Fabry-P\'erot resonator}
\author[1,2,*]{Yiqing Xu}
\author[1,2]{Stuart G. Murdoch}
\author[1,2]{St\'ephane Coen}
\author[1,2]{Miro Erkintalo}
\affil[1]{Department of Physics, University of Auckland, Auckland 1010, New Zealand}
\affil[2]{The Dodd-Walls Centre for Photonic and Quantum Technologies, Auckland 1010, New Zealand}
\affil[*]{Corresponding author: yxu079@aucklanduni.ac.nz}

\begin{abstract}
Dissipative Raman solitons in passive Kerr resonators have emerged as a promising route to broadband coherent frequency comb generation. Yet, their centre frequency has so far been mostly fixed near the Raman gain peak (13~THz downshifted from the pump centre frequency in silica-based fibres), constraining spectral coverage and compatibility with standard optical amplifiers. This limitation arises because the frequency shift of Raman solitons that fulfills phase-matching and group-velocity-matching conditions has to fall within the Raman gain band, leaving little room for spectral tuning when using a single conventional optical fibre. Here, we demonstrate that dispersion management of the fibre Fabry-P\'erot resonator which allows us to directly shift the soliton centre frequency. By combining two fibres with complementary dispersion profiles, we tailor the resonator's average dispersion to satisfy the phase-matching conditions for soliton formation at a target frequency downshifted by 7.8~THz from a pulsed pump, which is well outside the conventional 13~THz Raman gain band. This allows us to optically amplify the dissipative Raman soliton with a commercial L-band erbium-doped fibre amplifier, and fully characterise its temporal profile via the frequency-resolved optical gating technique.
\end{abstract}

\setboolean{displaycopyright}{true}

\begin{document}

\maketitle

\noindent The generation of coherent Kerr optical frequency combs can be linked to the formation of dissipative cavity solitons inside Kerr optical resonators~\cite{DelHaye2011,Herr2013,Kippenberg2018,Pasquazi2018}. Recent advances in high-finesse microresonators have enabled the nonlinear generation of cavity solitons with extremely low optical pump power~\cite{Yi2015,Yang2017,Li2022}. Meanwhile, broadband frequency combs in microresonators have been predominantly realised through resonator-dispersion engineering and spectral extension, allowing the comb bandwidth to span more than an octave~\cite{Moille2021,Song2024}. In recent years, additional approaches based on various gain processes have been developed to access different wavelength regions~\cite{Englebert2021,Moille2024,Liu2025,Xu2025}. Among these, stimulated Raman scattering is a well-known nonlinear gain mechanism that provides frequency-downshifted amplification relative to any strong pump laser source~\cite{Agrawal2001,Yang2017}. Leveraging this flexible nonlinear gain, recent works have demonstrated the generation of broadband coherent frequency combs, through the formation of dissipative Raman solitons in fibre resonators pulsed-driven at telecommunication band~\cite{Xu2021,Li2024}. However, the output energy from resonators is typically and calling for optical amplification outside of resonators, which poses a challenge for existing implementations. Because the wavelengths of these demonstrated Raman soliton combs mostly fall outside the operating range of commercial optical fibre amplifiers, or can lead to signal distortion when using semiconductor optical amplifiers~\cite{Sobhanan2022}, thereby restricting their broader applications.

In this letter, we overcome this wavelength limitation by managing the overall dispersion of a fibre resonator, and precisely fulfil the phase-matching and group-velocity-matching conditions required for dissipative Raman soliton formation at a target frequency~\cite{Li2024}. Concretely, we implement a dispersion-managed fibre cavity by combining two optical fibres with complementary dispersion characteristics, allowing the average resonator dispersion to be tailored with a degree of freedom unavailable in a single optical fibre. The target spectral window is the telecommunication L-band, chosen so that the generated dissipative Raman solitons can be directly amplified using a commercial L-band erbium-doped fibre amplifier (L-EDFA)--a requirement that prior work could not reliably satisfy. Driven by a pulsed pump centred at 1540~nm, the dispersion-engineered resonator produces dissipative Raman solitons at a centre wavelength of approximately 1600~nm. The temporal profile of the amplified solitons is subsequently characterised using frequency-resolved optical gating (FROG), providing direct validation that the generated pulses are consistent with the intended dissipative Raman soliton character.

We first recall the phase-matching conditions for dissipative Raman solitons under pulsed driving. The frequency shift of the generated dissipative Raman solitons is governed by the simultaneous linear phase-matching and group-velocity-matching conditions as~\cite{Li2024}
\begin{equation}\label{eqn:RS_PM}
    \Delta t/L \cdot\Omega_{\rm R} + \hat{D}(\Omega_{\rm R}) + q = 0,
\end{equation}
\begin{equation}\label{eqn:RS_GVM}
    \hspace{-11mm} \Delta t/L + \hat{D}_1(\Omega_{\rm R}) = 0,
\end{equation}
where $\Omega_{\rm R}/2\pi = (\omega_{\rm R} - \omega_{\rm p})/2\pi $ is the frequency shift of the dissipative Raman soliton centred at $\omega_{\rm R}/2\pi$ with respect to the pump frequency $\omega_{\rm p}/2\pi$, $L$ is the round-trip length of the fibre resonator and $q$ is the nonlinear phase shift per length when the Raman soliton is formed~\cite{Li2024}. $\hat{D}(\Omega)=\sum_{k\geq2}\beta_k\Omega^k/k!$ is the reduced dispersion of the fibre resonator, where $\beta_k$ is the $k$th-order dispersion coefficient obtained from the Taylor expansion of the resonator dispersion at $\omega_{\rm p}$. $\hat{D}_1(\Omega)=d\hat{D}/d\Omega$ represents the group-velocity mismatch with respect to the pump. $\Delta t = t_{\rm R}(\omega_{\rm p})-t_{\rm p}$ is the temporal desynchronisation between the driving pulse train period $t_{\rm p}$ and the round-trip time of the fibre resonator at the pump frequency $t_{\rm R}(\omega_{\rm p})$. By combining \eqref{eqn:RS_PM} and \eqref{eqn:RS_GVM} and neglecting the nonlinear phase shift $q$, the frequency shift of the dissipative Raman soliton $\Omega_{\rm R}/2\pi$ can be obtained by searching the roots of the polynomial
\begin{equation}\label{eqn:RS_Omega}
    \sum_{k\geq2}\beta_k\Omega_{\rm R}^k(k-1)/k!=0,
\end{equation}
and the corresponding temporal desynchronisation is determined by substituting $\Omega_{\rm R}$ back into \eqref{eqn:RS_PM}.

\begin{figure}[!htbp]
\centering
  \includegraphics[width = .95\linewidth, clip = true]{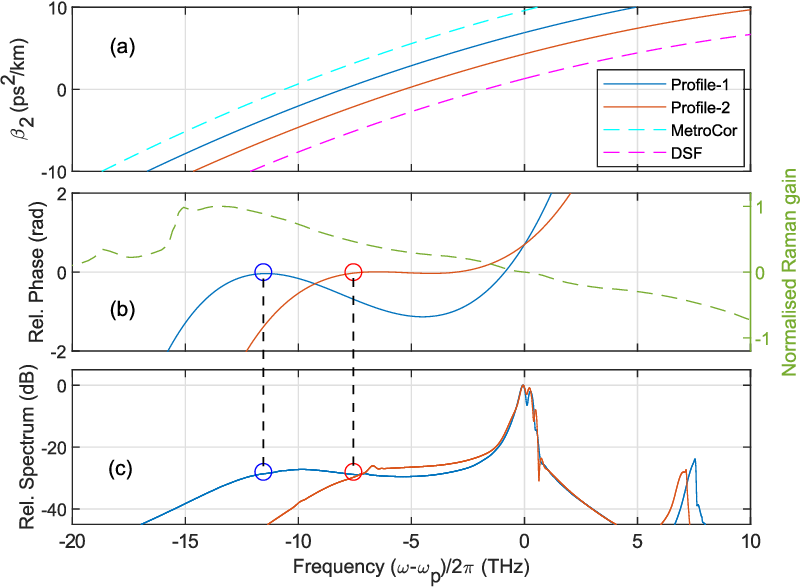}
  \caption{(a)~Second order dispersion $\beta_2$ of Corning MetroCor, DSF, Profile-1 and Profile-2. (b)~Phase mismatch calculated from reduced dispersion $\hat{D}(\omega-\omega_{\rm p})$, and (c)~simulated output spectra, using the dispersion-managed fibre parameters of Profile-1 and Profile-2 in Table.~\ref{tab:sim_parameters}. Green dashed curve in (b) is the normalised Raman gain spectrum. Circles indicate the phase-matching frequency of the dissipative Raman solitons.}\label{Fig:fig1}
\end{figure}

\vspace{-4mm}
\begin{table}[!htbp]
\centering
\tiny
\caption{\bf Parameters for dispersion management of fibre resonator using Corning MetroCor and DSF at 1540~nm.}
\resizebox{.4\textwidth}{!}{%
\begin{tabular}{rrr}
\hline
     & Profile-1 & Profile-2 \\
\hline
    Fiber & MetroCor+DSF & MetroCor+DSF \\
    2$\times$~Length~(m) & 0.25+0.12 & 0.25+0.44 \\
    $\Delta t/L$~(fs/m) & 146 & 71 \\
    $\overline{\beta}_2$~(ps$^2$/km) & 6.9 & 4.3 \\
    $\beta_3$~(s$^3$/m) & \multicolumn{2}{c}{ \hspace{7.1mm} $0.11\times10^{-39}$} \\
    $\beta_4$~(s$^4$/m) & \multicolumn{2}{c}{ \hspace{8mm} $-9\times10^{-55}$} \\
    $\Omega_{\rm R}/2\pi$~(THz) & $-11.9$ &  $-7.8$\\
\hline
\end{tabular}
}
  \label{tab:sim_parameters}
\end{table}
Practically, once the target frequency of the dissipative Raman soliton $\omega_{\rm R}/2\pi$ (e.g., centred within the L-EDFA in our case) and the pump frequency are specified, the dispersion profile that satisfies the phase-matching condition can be determined. To demonstrate how to obtain the target phase-matching dispersion profile, we consider mainly adjusting the second-order dispersion coefficient of the fibre resonator through dispersion management~\cite{Jang2014,Luo2015}, whilst assuming the higher-order dispersion parameters of the individual fibre segments as approximately equal for simplicity. The averaged second-order dispersion of the resonator is given by $\bar{\beta}_2 = \sum{\beta_{2,n} L_n}/\sum{L_n}$, where $n$ denotes each fibre segment. For a desired frequency shift $\Omega_{\rm R}/2\pi$ of the dissipative Raman soliton (within the Raman gain bandwidth), the averaged dispersion that satisfies the phase-matching condition can be flexibly set by adjusting the relative lengths of the constituent fibre segments. To illustrate this dispersion-profile engineering via fibre-length management, we consider a resonator constructed by directly splicing two commercial optical fibres: Corning MetroCor fibre and Corning dispersion-shifted fibre (DSF) with the second-order dispersion coefficients $\beta_{\rm 2,MetroCor} =9.6$~ps$^2$/km and  $\beta_{\rm 2,DSF}=1.3$~ps$^2$/km, respectively, both evaluated at the 1540~nm. The lengths and dispersion coefficients of higher-order for both fibres are listed in Table~\ref{tab:sim_parameters}. Figure~\ref{Fig:fig1}~(a) showcases the second-order dispersion profiles of the MetroCor fibre and the DSF, alongside two representative dispersion-managed profiles (Profile-1 and Profile-2), demonstrating how dispersion management shifts the phase-matching condition and thereby controls the centre frequencies of the dissipative Raman solitons. The corresponding phase-matching conditions for Profile-1 and Profile-2, evaluated using \eqref{eqn:RS_PM}, are shown in Fig.~\ref{Fig:fig1}~(b). The desynchronisation parameters $\Delta t/L$ associated with each profile are calculated from \eqref{eqn:RS_GVM} and summarised in Table~\ref{tab:sim_parameters}, as well as the corresponding centre frequencies of the dissipative Raman solitons by solving \eqref{eqn:RS_Omega}.

Since the fibre resonator is dispersion managed, the periodic modulation of the dispersion profile arising from different fibre segments will inevitably induce Kelly sidebands during optical field propagation~\cite{Luo2015}. To fully obtain the spectral components of dissipative Raman solitons, we simulate the intracavity field inside a fibre Fabry-P\'erot resonator using an Ikeda-like map based on the nonlinear Schrödinger equation. The evolution of the optical field at the $m$-th round trip is described as~\cite{Ikeda1979,Moille2024}
\begin{equation}\label{eqn:RamanIkeda}
\begin{split}
    \frac{\partial E^{(m)}(z,\tau)}{\partial z} & = \left[-\frac{\Delta t}{L}\frac{\partial}{\partial\tau} + i\sum_{k\geq2}\frac{\beta_k}{k!}\left(i\frac{\partial}{\partial\tau}\right)^k\right] E^{(m)}\\
    & + i\gamma\left[(1-f_{\rm R})|E^{(m)}|^2 + f_{\rm R} h_{\rm R}*|E^{(m)}|^2\right]E^{(m)},
\end{split}
\end{equation}
where $E(z,\tau)$ is the slowly varying envelope of the intracavity field, $z$ is the propagation distance along the optical fibre, and $\tau$ denotes the time coordinate in a reference frame moving at the group-velocity corresponding to the coherently driving field frequency $\omega_{\rm p}$. Here, $\gamma$ is the Kerr nonlinear coefficient, $f_{\rm R}=0.18$ is the Raman fraction, and $h_{\rm R}$ is the Raman response function in the time domain~\cite{Agrawal2001}. The Ikeda-like map is completed by imposing a boundary condition that accounts for the coupling of the external pulsed driving field into the cavity as
\begin{equation}\label{eqn:Ikeda_boundary}
    E^{(m+1)}(0,\tau) = \rho E^{(m)}(L,\tau)\exp(-i\delta_0) + \sqrt{\theta}E_{\rm p}(\tau),
\end{equation}
where $\rho$ represents the round-trip power loss of the fibre resonator, $\theta$ is the input coupling transmission, and $\delta_0=2\pi k-\beta(\omega_{\rm p})L$ denotes the detuning of the pump from the nearest resonance. In the simulations, the input pump field has a Gaussian temporal profile $E_{\rm p}(\tau)=\sqrt{P}\exp(-\tau^2/\tau^2_0)$, where $\tau_0=\tau_{\rm 3dB}/\sqrt{2\ln2}$ is related to the full-width half-maximum temporal width $\tau_{\rm 3dB}$. We consider a fibre resonator which has the dispersion profiles in Table.~\ref{tab:sim_parameters}, with a nonlinear coefficient of $\gamma=2$~W$^{-1}$/km. The round-trip power transmission is set to $\rho=0.95$, and the input coupling transmission is $\theta=0.05$. By pumping the fibre resonator with a pulse width $\tau_{\rm 3dB}=1.8$~ps and peak power $P=28$~W at a cavity detuning of $\delta_0=0.18$, we simulate the intracavity field evolution over 1000 round trips. The resulting output spectra for each dispersion-management profile are plotted in Fig.~\ref{Fig:fig1}~(c). As expected, the frequency centres of the dissipative Raman solitons are in excellent agreement with the phase-matching frequencies predicted in Fig.~\ref{Fig:fig1}~(b), particularly with the centre frequency of Profile-2 Raman soliton located inside L-EDFA amplification band. As shown by the Raman gain spectrum (green dashed curve) in Fig.~\ref{Fig:fig1}(b), even though the dissipative Raman solitons experience reduced gain as their frequency shift moves away from the Raman gain peak, the phase-matching condition primarily determines the frequency shift at which dissipative Raman solitons can stably phase-lock to the pulsed driving field~\cite{Li2024}.
\begin{figure}[!htbp]
\centering
    \vspace{-2mm}
    \includegraphics[width = .9\linewidth, clip = true]{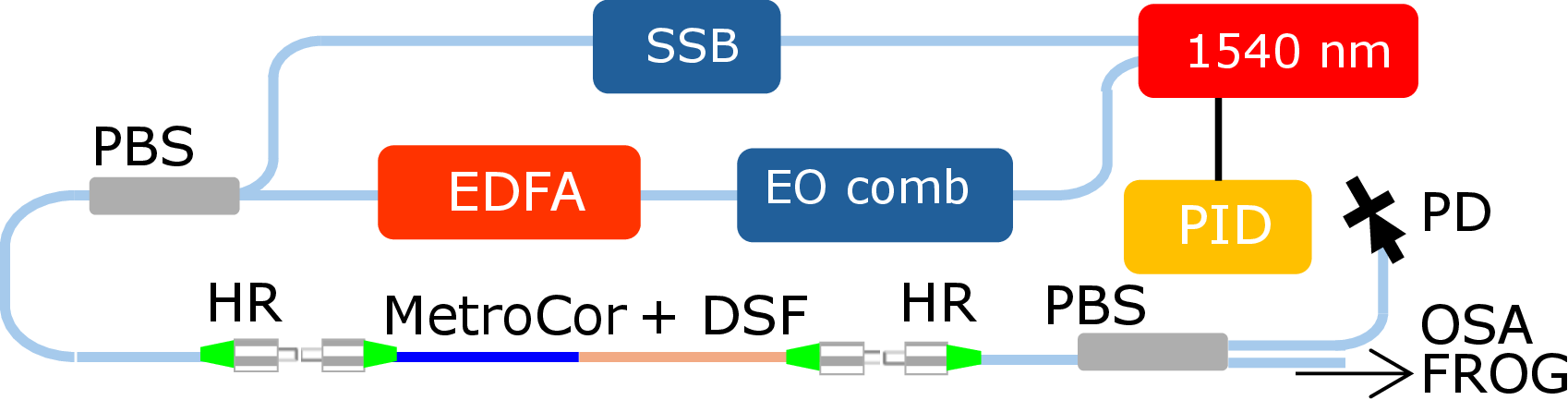}
    \caption{Experimental schematic. SSB:~single sideband modulator, EDFA:~erbium doped fibre amplifier, PBS:~polarisation beam splitter, HR:~high reflector.}\label{Fig:fig2a}
\end{figure}

To realise the dispersion-engineered resonator predicted above that supports dissipative Raman soliton formation in the telecommunication L-EDFA, we construct the experimental setup schematically shown in Fig.~\ref{Fig:fig2a}. The resonator is built around a Fabry-P\'erot fibre cavity by butt-coupling two high reflectors (HRs) to the ends of a dispersion-managed fibre, whose parameters are chosen to match Profile-2 in the numerical simulations above. The splicing loss between the two fibre segments is measured to be approximately $\sim5$\%. At the input end, a $\sim95$\% HR is used to achieve approximately critical coupling by matching the round-trip cavity loss, while the output end is coupled to a $\sim$99\% HR for monitoring the intracavity field. As a result, the actual round-trip loss and input transmission closely match the parameters used in the simulations, and the overall finesse of the fibre resonator is measured to be 60. The pulsed pump for coherently driving our fibre resonator is derived from a narrow-linewidth continuous-wave seed laser operating at 1540~nm (NKT, Koheras-BASIK). The seed laser is converted into a pulse train using an electro-optic~(EO) comb generator comprising a phase modulator, an intensity modulator, and a sequence of linear and nonlinear fibre-based pulse compression stages~\cite{Xu2020}. The resulting pump pulses have a temporal width of 1.8~ps. The repetition rate of the pump pulse train is precisely adjusted to match the round-trip time of the fibre resonator. Fine control of the temporal desynchronisation is achieved by tuning the radio-frequency clock (not shown) that drives the EO comb. After generation, the pump pulses are amplified using a C-band EDFA and subsequently coupled into the resonator through a polarising beam splitter (PBS). In addition to the pump field, a frequency-shifted control beam is generated from the same seed laser and injected into the orthogonal polarisation of the fibre resonator via the second input port of the PBS. This control beam is employed for stabilisation of the cavity detuning. At the output end of the resonator, the pump and control beams are separated by a second PBS and independently monitored. The transmitted control beam is detected using a slow amplified photodetector~(PD), and the resulting signal is fed into a proportional–integral–derivative~ (PID) controller (SRS, SIM960). The feedback signal from the PID controller is applied to the seed laser to stabilise the cavity detuning~\cite{Nielsen2021}. The intracavity field in the pump polarisation is characterised using an optical spectrum analyser~(OSA) and a frequency-resolved optical gating~(FROG) system~\cite{Kane1993}.
\begin{figure}[!htbp]
\centering
  \includegraphics[width = .9\linewidth, clip = true]{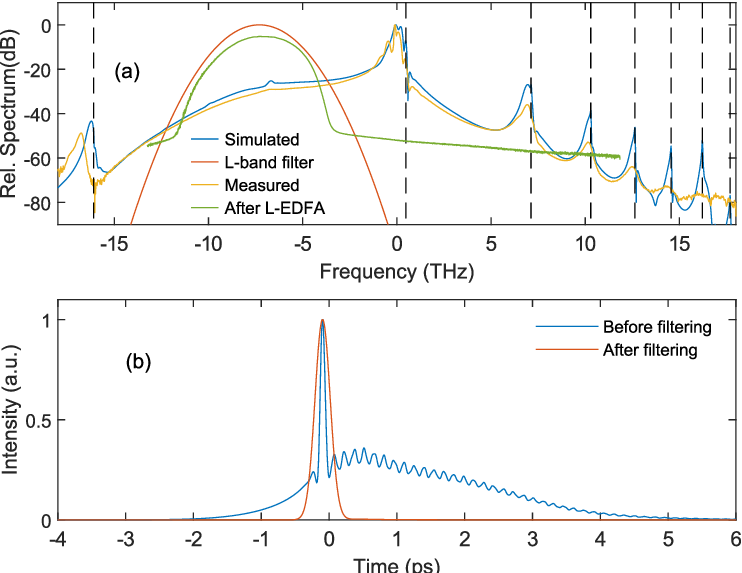}
  \caption{(a)~Simulated (blue) and experimental (yellow) output spectra using dispersion management Profile-2. Black vertical dashed lines indicate the phase-matching frequency of the Kelly sidebands. Green curve is measured spectrum of the dissipative Raman soliton after amplified by a low-noise L-EDFA. Red curve is the numerical L-band spectral filter. (b)~Simulated temporal profile of dissipative Raman soliton (blue) based on dispersion management Profile-2, and the temporal profile (red) after spectrally filtered by L-band filter in (a).}\label{Fig:fig2}
\end{figure}

After stabilising the cavity, we first observe the generation of a dissipative Raman soliton and plot the experimentally measured output spectrum (yellow curve) in Fig.~\ref{Fig:fig2}~(a). For comparison, the corresponding simulated output spectrum (blue curve) from Fig.~\ref{Fig:fig1}~(c) is also plotted atop, showing good agreement with the experimental measurement, having the centre frequency downshifted by $7.8$~THz ($\sim1604$~nm). Owing to the periodic dispersion profile experienced by the optical field when resonating, Kelly sidebands are generated at frequencies that accumulate a phase shift over one round trip equal to $2\pi m$ ($m$ is an integer). These frequencies can be accordingly predicted by $\Delta t\cdot\Omega + \hat{D}(\Omega)L + \delta_0 = 2\pi m$~\cite{Luo2015}. We calculate the phase-matching frequencies of the Kelly sidebands and overlay them on Fig.~\ref{Fig:fig2}~(a), indicated by the black dashed vertical lines. These predicted phase-matching frequencies coincide well with the sub-peaks observed in both the simulated and experimental spectra. The experimental results shown in Fig.~\ref{Fig:fig2} confirm (i) that Raman solitons can exist in dispersion managed resonators and (ii) that dispersion management allows the solitons to be generated within the L-EDFA spectral windows. The ability to amplify the Raman solitons makes it possible for the first time to characterise their temporal features. We then spectrally filter the output field by removing wavelength components below 1570~nm using a wavelength-division multiplexor to ensure the L-EDFA only amplifies the spectral signal within its operational wavelength region. The remaining spectral components are subsequently amplified using a low-noise L-EDFA. The green curve in Fig.~\ref{Fig:fig2}~(a) shows the measured spectrum of the amplified dissipative Raman soliton. Note that since the output power of the dissipative Raman solitons is relatively low, the nonlinear effects occurring within the L-EDFA are negligible.

\begin{figure}[!htbp]
\centering
  \includegraphics[width = .9\linewidth, clip = true]{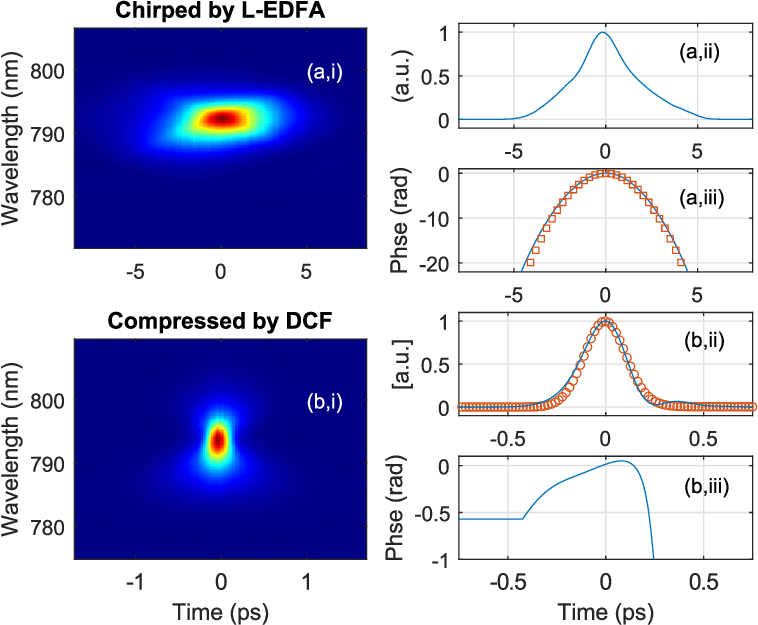}
  \caption{(a,i) and (b,i)~Experimental FROG measurements of dissipative Raman soliton before and after the DCF compression. (a,ii) and (b,ii)~Corresponding recovered temporal intensity profiles, (a,iii) and (b,iii)~recovered temporal phase profiles. Squares in (a,iii) are the simulated the temporal phase before the DCF fibre, and circles in (b,ii) are the simulated temporal intensity profile from Fig.~\ref{Fig:fig2}~(b) after spectral filtering.}\label{Fig:fig3}
\end{figure}
Before making the temporal measurement, we first numerically observe the temporal profile of this spectrally clipped Raman soliton due to the finite spectral gain of the L-EDFA. The spectral filtering is approximated by a Gaussian filter, shown as the red curve in Fig.~\ref{Fig:fig2}~(a). The corresponding simulated temporal profile of the unfiltered temporal output field is plotted as the blue curve in Fig.~\ref{Fig:fig2}~(b), together with the temporal profile obtained after applying the numerical spectral filter, expecting the temporal width of the amplified Raman soliton to be broaden into 230~fs. Following amplification, the temporal profile of the dissipative Raman soliton is characterised using the FROG. The retrieved FROG spectrogram, temporal intensity, and temporal phase are shown in Fig.~\ref{Fig:fig3}~(a,i–iii), respectively. The reconstructed temporal phase reveals that the amplified output field is strongly chirped. We attribute this chirp primarily to the accumulated dispersion arising from the long fibre sections and doped fibres within the L-EDFA. By approximating the temporal phase [shown as square markers in Fig.~\ref{Fig:fig3}~(a,iii)] with a quadratic profile, we estimate the amount of chirp induced by the L-EDFA. To recompress the chirped Raman soliton, we pass the field through a 3~m long Corning dispersion compensating fibre~(DCF, $\beta_2=140$~ps$^2$/km) whose length is calculated from the estimated chirp. After recompressing by the DCF, another FROG measurement is performed, as shown in Fig.~\ref{Fig:fig3}~(b,i). The retrieved temporal intensity and phase after recompression are plotted in Fig.~\ref{Fig:fig3}~(b,ii) and (b,iii), respectively. For comparison, the simulated temporal intensity profile from Fig.~\ref{Fig:fig2}~(b) is overlaid with the experimentally recovered temporal intensity. The excellent agreement between the two confirms that the recompressed dissipative Raman soliton is nearly transform-limited.

In conclusion, we have presented an experimental study on the generation of ultrashort dissipative Raman solitons in a dispersion-managed fibre resonator. In contrast to conventional dissipative Raman solitons, whose frequency centre typically coincides with the Raman gain peak, our dispersion-managed fibre resoantor enables to spectrally tailor of the soliton centre frequency to an on-demand value by simply combining two commercial optical fibre segments. Our simulations accurately predict the experimental output spectra. In particular, in this study we have engineered the dispersion such that the dissipative Raman solitons fall within the gain region of a commercial L-band erbium-doped fibre amplifier. Taking advantage of optical amplification, we have been able to fully characterise the temporal profile using the FROG technique and to recompress the amplified pulse close to its transform limit.

\begin{backmatter}

\bmsection{Funding} Marsden Fund of the Royal Society of New Zealand.

\bmsection{Disclosures} The authors declare no conflicts of interest.

\bmsection{Data availability} Data underlying the results presented in this paper are not publicly available at this time but may be obtained from the authors upon reasonable request.\\

\end{backmatter}


\begin{thebibliography}{10}
\newcommand{\enquote}[1]{}

\bibitem{DelHaye2011}
P.~Del’Haye, T.~Herr, E.~Gavartin, et~al., \enquote{Octave Spanning Tunable Frequency Comb from a Microresonator,} Phys. Rev. Lett. \textbf{107}, 063901 (2011).

\bibitem{Herr2013}
T.~Herr, V.~Brasch, J.~D.~Jost, et~al, \enquote{Temporal solitons in optical microresonators,} Nat. Photonics \textbf{8}, 145 (2013).

\bibitem{Kippenberg2018}
T.~J.~Kippenberg, A.~L.~Gaeta, M.~Lipson, and M.~L.~Gorodetsky, \enquote{Dissipative Kerr solitons in optical microresonators,"} Science \textbf{361}, eaan8083 (2018).

\bibitem{Pasquazi2018}
A.~Pasquazi, M.~Peccianti, L.~Razzari, et~al, \enquote{Micro-combs: A novel generation of optical sources,} Phys. Rep. \textbf{729}, 1-81 (2018).

\bibitem{Yang2017}
Q.~F.~Yang, X.~Yi, K.~Y.~Yang, and K.~Vahala, \enquote{Stokes solitons in optical microcavities,} Nat. Phys. \textbf{13}, 53-57 (2017).

\bibitem{Yi2015}
X.~Yi, Q.-F.~Yang, K. Y.~Yang, M.-G.~Suh, and K.~Vahala, \enquote{Soliton frequency comb at microwave rates in a high-Q silica microresonator,} Optica \textbf{2}, 1078-1085 (2015).

\bibitem{Li2022}
J.~Li, C.~Bao, Q.-X.~Ji, H.~Wang, L.~Wu, S.~Leifer, C.~Beichman, and K.~Vahala, \enquote{Efficiency of pulse pumped soliton microcombs,}" Optica \textbf{9}, 231-239 (2022).

\bibitem{Moille2021}
G.~Moille, E.~F.~Perez, J.~R.~Stone, et~al., \enquote{Ultra-broadband Kerr microcomb through soliton spectral translation,} Nat. Commun. \textbf{12}, 7275 (2021).


\bibitem{Song2024}
Y.~ Song, Y.~Hu, X.~Zhu, K.~Yang, and M.~Lončar, \enquote{Octave-spanning Kerr soliton frequency combs in dispersion- and dissipation-engineered lithium niobate microresonators,} Light: Sci. Appl. \textbf{13}, 225 (2024).

\bibitem{Englebert2021}
N.~Englebert, F.~De~Lucia, P.~Parra-Rivas, et~al., \enquote{Parametrically driven Kerr cavity solitons,} Nat. Photonics \textbf{15}, 857-861 (2021).


\bibitem{Liu2025}
P.~Liu, Q.-X.~Ji, J.-Y.~Liu, et~al., \enquote{Near-visible integrated soliton microcombs with detectable repetition rates,} Nat. Commun. \textbf{16}, 4780 (2025).

\bibitem{Xu2025}
Y.~Xu, S.~Coen, M.~Erkintalo, and S.~G.~Murdoch, \enquote{Toward visible ultrafast imaging with a synchronously pumped switching wave Kerr frequency comb,} Opt. Express \textbf{33}, 4714-4724 (2025).

\bibitem{Moille2024}
G.~Moille, M.~Leonhardt, D.~Paligora, et~al., \enquote{Parametrically driven pure-Kerr temporal solitons in a chip-integrated microcavity,} Nat. Photonics \textbf{18}, 617-624 (2024).


\bibitem{Agrawal2001}
G.~P.~Agrawal, \enquote{Nonlinear fiber optics,} 3rd ed., Optics and photonics (Academic Press, San Diego, 2001)


\bibitem{Xu2021}
Y.~ Xu, A.~Sharples, J.~Fatome, et~al., \enquote{Frequency comb generation in a pulse-pumped normal dispersion Kerr mini-resonator,} Opt. Lett. \textbf{46}, 512-515 (2021).


\bibitem{Li2024}
Z.~Li, Y.~Xu, S.~Shamailov, et~al, \enquote{Ultrashort dissipative Raman solitons in Kerr resonators driven with phase-coherent optical pulses,} Nat. Photonics \textbf{18}, 46-53 (2024).

\bibitem{Sobhanan2022}
A.~Sobhanan, A.~Anthur, S.~O’Duill, et~al., \enquote{Semiconductor optical amplifiers: recent advances and applications,} Adv. Opt. Photon. \textbf{14}, 571-651 (2022)


\bibitem{Jang2014}
J.~K.~Jang, M.~Erkintalo, S.~G.~Murdoch, and S.~Coen, \enquote{Observation of dispersive wave emission by temporal cavity solitons,} Opt. Lett. \textbf{39}, 5503-5506 (2014).

\bibitem{Luo2015}
K.~Luo, Y.~Xu, M.~Erkintalo, and S.~G.~Murdoch, \enquote{Resonant radiation in synchronously pumped passive Kerr cavities,} Opt. Lett. \textbf{40}, 427-430 (2015).


\bibitem{Ikeda1979}
K.~Ikeda, \enquote{Multiple-Valued Stationary State and Its Instability of the Transmitted Light by a Ring Cavity System,} Optics Commun. \textbf{30}, 257-261 (1979).

\bibitem{Xu2020}
Y.~Xu, Y.~Lin, A.~Nielsen, et~al., \enquote{Harmonic and rational harmonic driving of microresonator soliton frequency combs,} Optica \textbf{7}, 940-946 (2020).

\bibitem{Nielsen2021}
A.~U.~Nielsen, Y.~Xu, C.~Todd, et.~al, \enquote{Nonlinear Localization of Dissipative Modulation Instability,} Phys. Rev. Lett. \textbf{127}, 123901 (2021).

\bibitem{Kane1993}
D.~J.~Kane, and R.~Trebino, \enquote{Characterization of arbitrary femtosecond pulses using frequency-resolved optical gating,} IEEE Journal of Quantum Electronics \textbf{29}, 571-579 (1993).


\end{thebibliography}
\end{document}